\documentclass[12pt,letterpaper]{article}
\usepackage[utf8]{inputenc}

\usepackage[letterpaper,bindingoffset=0.1in,%
            left=1in,right=1in,top=1in,bottom=1in,%
            footskip=0.25in]{geometry}

\usepackage{setspace} 
\usepackage{mathptmx} 
\usepackage{xcolor}  
\usepackage{dsfont}  
\usepackage[hidelinks]{hyperref} 

\usepackage{tikz} 
\usetikzlibrary{positioning, shapes,arrows, calc, arrows.meta} 

\usepackage{float} 
\usepackage{graphicx} 
\usepackage{subcaption} 
\usepackage[labelfont=bf]{caption} 
\usepackage{enumitem} 
\usepackage{amstext} 
\usepackage{amsmath}
\usepackage{array,multirow} 
\usepackage[bottom]{footmisc} 
\usepackage{url}

\usepackage[pagewise]{lineno} 
\usepackage{authblk}  
\usepackage{dsfont}  
\usepackage{amsfonts} 
\usepackage[round]{natbib}
\hypersetup{colorlinks, citecolor=black, linkcolor=black, urlcolor=blue}
\usepackage{hyperref}
\newcolumntype{K}[1]{>{\centering\arraybackslash}p{#1}}  

\begin{document}

\title{\textbf{Distribution of Gaps in Multi-lane Orderly and Disorderly Traffic Streams}}
\date{}
\author[1]{Ankita Sharma}
\author[1]{Partha Chakroborty}
\author[1]{Pranamesh Chakraborty}
\affil[1]{Department of Civil Engineering, Indian Institute of Technology Kanpur, India}
\maketitle

\begin{abstract}
To study gap acceptance behaviour one needs the distribution (or probability density function) of gaps in the opposing stream. Further, in these times of widespread availability of large computing powers, traffic simulation has emerged as a popular analysis and design tool. Such simulations rely on randomly generating the arriving vehicles in a way that statistically resembles real-world streams. The generation process for disorderly streams requires information on gap distributions. A study of past literature reveals that very little work has been done to determine the distribution of gaps on multi-lane orderly and disorderly streams. This study aims to develop an analytical framework to specify the distribution of gaps for such streams. This analytical framework is built using the Renewal Process Theory. A maximum likelihood based process for the estimation of the parameters of the analytically derived distribution is also described. Later, real-world gap data from three different sites covering orderly and disorderly streams are used to show how the derived distribution function (using the proposed method) ably describes the observed gap distributions.

\end{abstract}

\section{Introduction}
Gap acceptance is a basic driving task. In this, a driver evaluates gaps in an opposing stream (which he/she wants to cross) and accepts one that is deemed adequate. Three basic questions emerge: (i) how does one define a gap in an opposing (or any) stream that is either a single-lane or a multi-lane orderly stream or a disorderly stream on a wide road? (ii) do drivers evaluate and react to physical gaps or their perceived values? and (iii) how does one model/mathematically describe the gap acceptance process? While many attempts have been made in the past to understand the last question (for example, \citealt{tanner1951}, \citealt{miller1971nine}, \citealt{RengarajuRao1995}, \citealt{TIAN1999187} and others) the first two questions have remained largely unanswered. The goal of the present work is to investigate the issues arising out of the first question.

In the gap acceptance literature, gaps are generally defined as the time interval between arrivals of two successive vehicles (\textcolor{black}{\citealt{garwood1940}, \citealt{tanner1951}, \citealt{mayne1954}, \citealt{ashworth1968note}, \citealt{Daganzo1981}}). Therefore, for single-lane orderly streams, a gap is the same as a headway. Hence, a headway distribution can be used to describe the distribution of gaps (for example, see \citealt{tanner1951}, \citealt{mayne1954}). However, the answer is not as straightforward for multi-lane orderly streams. In this case, there has to be a simultaneous absence of vehicles on all the lanes for a duration that can be considered as a gap. That is, a gap is not identical to a headway in any of the lanes. \citet{TIAN1999187} talks about this but does not investigate further. Others, like \citet{tanner1951} who work with gap acceptance in multi-lane orderly streams simply assume gaps to be distributed exponentially like one would do when dealing with single-lane traffic. Yet others, like \textcolor{black}{\citet{garwood1940}, \citet{ashworth1968note}, \citet{miller1971nine}}, continue to assume an exponential distribution without any discussion on how many lanes are present in their opposing stream.

In disorderly streams, where lane discipline is weak at best, only gaps can be defined; headways in the classical sense cannot be defined. Interestingly, much like the case in multi-lane orderly streams, past researchers, without any stated justification, typically assume gaps on such streams to be exponentially distributed (for example, see \citealt{RengarajuRao1995}). While others like \citet{patilandpawar}, and \citet{dutta_ahmed} try to fit various theoretical distributions to observe gaps.

It is clear that little or no attention has been paid to the determination of the distribution of gaps in multi-lane orderly streams or disorderly streams. Yet, understanding the distribution is paramount while analysing properties of queues that develop in the stream of vehicles waiting to cross. For example, determination of distributions and moments of waiting times and queue lengths all require information on gap distribution. Design ramifications include decisions on auxiliary lane lengths, signalization of intersections etc. The importance of gap distributions is even greater for disorderly streams since the simulation of such streams without an understanding of gap distribution is akin to simulating orderly streams without information on headway distribution. Surprisingly, amongst the many simulation studies of disorderly streams the authors referred to (for example, \citealt{SreekumarandMathew}, \citealt{matchaetal}, \citealt{Azametal}) none discuss, in detail, the subject of vehicle generation which would require a discussion on time gaps between successive arrival of vehicles. Some like \citet{HossainandMcDonald}, \citet{ArasanandKoshy}, \citet{Deyetal} simply assume a distribution and generate the vehicles.

Given the reliance of traffic design and related what-if studies on computer simulations (more so for disorderly streams) a detailed study, both analytical and empirical, on gap distributions must be undertaken for both orderly and disorderly streams. This is the objective of the present work. The next section elaborates on how gaps can be defined in multi-lane orderly and disorderly streams. Section 3 presents an analytical formulation to determine the probability density function (pdf) of gaps. Section 4 describes the data on multi-lane orderly streams and disorderly streams used in this study to explore the applicability of the analytically obtained pdf of gaps. Section 5 dwells on how the parameters of the pdf can be estimated from real-world data; the results from this estimation process are presented in the next section. The last section (Section 7) concludes the paper with a discussion of its major findings.

\section{More on Defining Gaps} \label{prob_stat}
\begin{figure}[h!]
\begin{center}
    \includegraphics[width=0.7\textwidth]{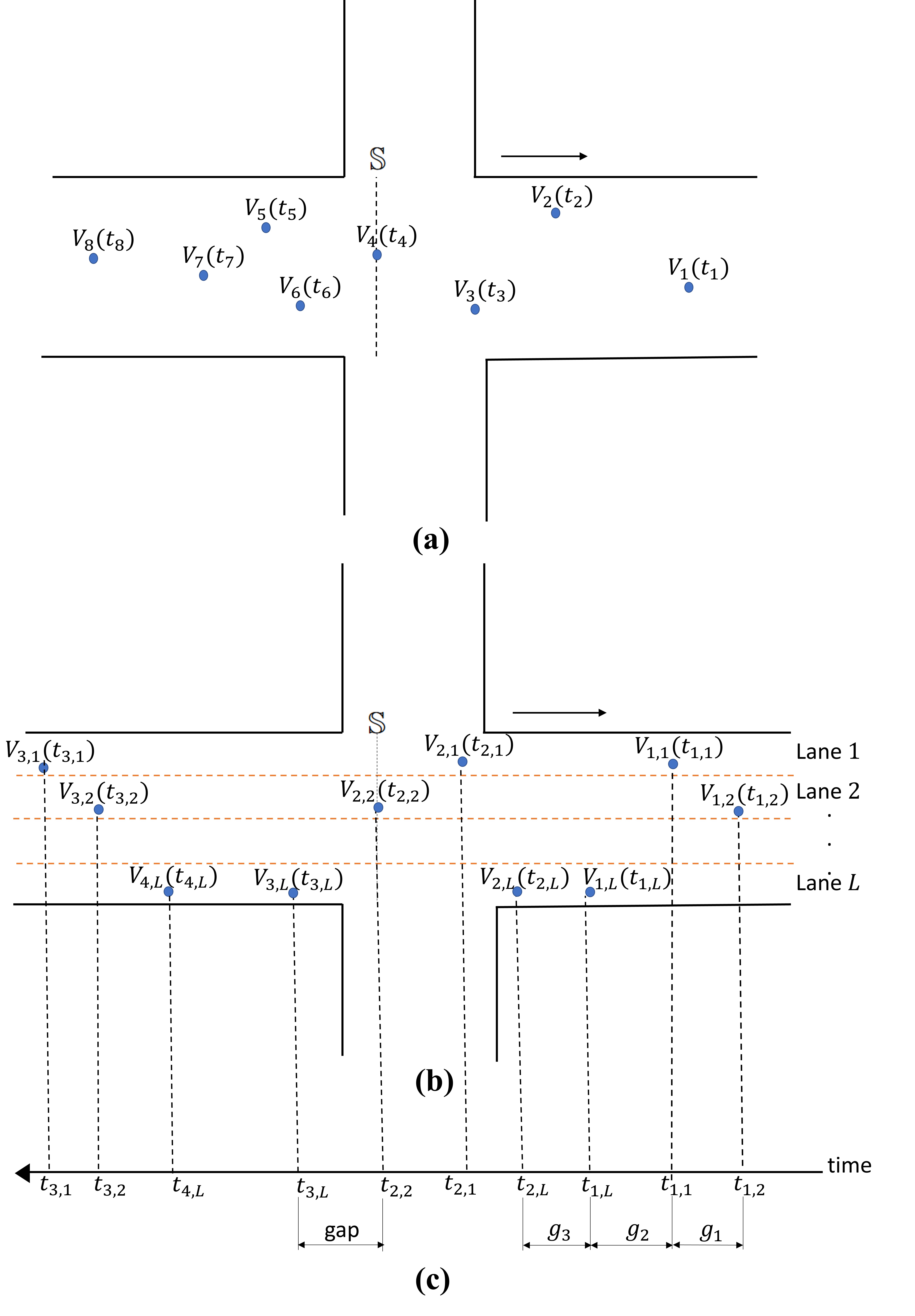}
    \caption{\label{fig:gap_fig1}Snapshot of a road section (a) at time $t_4$ in disorderly stream, (b) at time $t_{2,2}$ in orderly stream and (c) time ($t_{i,j}$), when Vehicle $i$ in Lane $j$ crosses Section $\mathbb{S}$. plotted on time axis for orderly stream.}
\end{center}    
\end{figure}

A gap is defined as the time duration between two successive arrivals at a given section of a road (see, \citealt{garwood1940}, \citealt{tanner1951}, \citealt{mayne1954}, \citealt{ashworth1968note}). The position of the section is only defined longitudinally and hence the lateral locations of vehicles (as they cross the section) do not play a role in the determination of gaps. For example in Figure \ref{fig:gap_fig1}(a) that represents a disorderly stream or Figure \ref{fig:gap_fig1}(b) that represents a multi-lane orderly stream all that matters in the determination of gaps at Section $\mathbb{S}$ are the times when vehicles $V_1, \ V_2, \ V_3,...$ (where $V_i$ is the $i^{\text{th}}$ vehicle) in Figure \ref{fig:gap_fig1}(a) and $V_{1,1},\ V_{1,2},\  V_{1,3},...,\ V_{2,1},\ V_{2,2}, \ V_{2,3}, ...$ (where $V_{i,j}$ is the $i^{\text{th}}$ vehicle on Lane $j$) in Figure \ref{fig:gap_fig1}(b) cross Section $\mathbb{S}$ (Note that the lines extending from Figure \ref{fig:gap_fig1}(b) will be used in Figure \ref{fig:gap_fig1}(c) which will be introduced later). It does not matter that, for example, $V_1,\ V_2,\ V_3,...$ are in different lateral locations in Figure \ref{fig:gap_fig1}(a) or $V_{1,1}$ and $V_{1,2}$, etc, are in different lanes in Figure \ref{fig:gap_fig1}(b). 

In the figure, the times when vehicles cross Section $\mathbb{S}$ are mentioned parenthetically. For the present, to keep things streamlined, the discussions will be for orderly streams. Extensions for disorderly streams will be introduced later. 

As shown in the Figure \ref{fig:gap_fig1}(b), $t_{i,j}$ is the time when $V_{i,j}$ crosses Section $\mathbb{S}$. Imagine these $t_{i,j}$ being plotted on the time axis as shown in Figure \ref{fig:gap_fig1}(c). Then a gap is simply the time difference between two successive points on the time axis plot shown in \ref{fig:gap_fig1}(c)\footnote{The gap (or for that matter headway) cannot be identified on a figure like Figure \ref{fig:gap_fig1}(b) because they do not represent the time dimension.}. Note, the plot in Figure \ref{fig:gap_fig1}(c) is nothing but an arrangement of $t_{i,j}$ in ascending order and a gap is the difference of two successive time values in this ordered list. The first few gaps $g_1$, $g_2$ and $g_3$ are also marked in Figure \ref{fig:gap_fig1}(c).

For disorderly streams, one would proceed the same way. Except now, one would arrange $t_{i}$ of Figure \ref{fig:gap_fig1}(a) in ascending order. The reason why there is no distinction between the definition of gaps in orderly and disorderly streams is because these streams differ only in their characteristic in the spatial dimension and gaps are a property in the temporal dimension. 

The next section presents a mathematical formulation that can be employed to determine the cumulative distribution function (cdf) and probability density function (pdf)  of gaps for both orderly and disorderly streams. Later, the maximum likelihood estimation (MLE) technique is used to estimate the parameters of the density function. Open source data on multi-lane orderly streams and collected data on disorderly streams are studied to see the effectiveness of the current formulation.  

\section{Determination of pdf of Gap} \label{sec:dist_gap}
The traditional representation of gap as introduced in the previous section while instructional and easy to comprehend cannot be readily used to determine its pdf. In this section, a different representation will be employed to evaluate the gaps. Note, that while the representation will be different from that introduced in the previous section, the results (that is, the values of the gaps) will, of course, be the same. It is this new representation of gaps that will be used to derive the cdf and pdf of gaps, $G$. As before, the discussion will assume orderly streams to begin with. Later, the results will be extended to disorderly streams. 
\begin{figure}[h!]
\begin{center}    
\begin{subfigure}{\textwidth}
  \centering
  \includegraphics[width=0.6\textwidth]{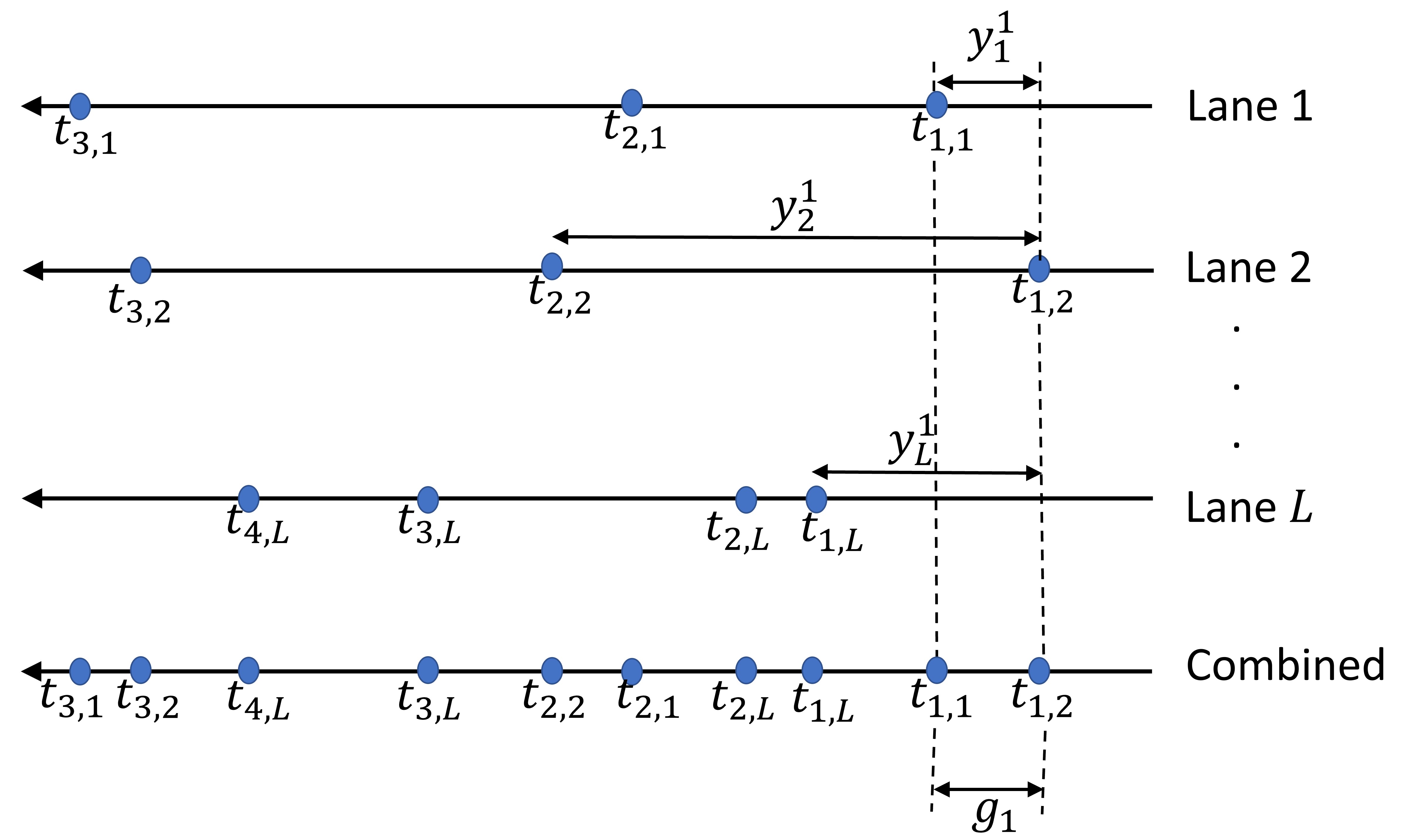}
  \caption{\label{g1}}
\end{subfigure}
\begin{subfigure}{\textwidth}
  \centering
  \includegraphics[width=0.6\textwidth]{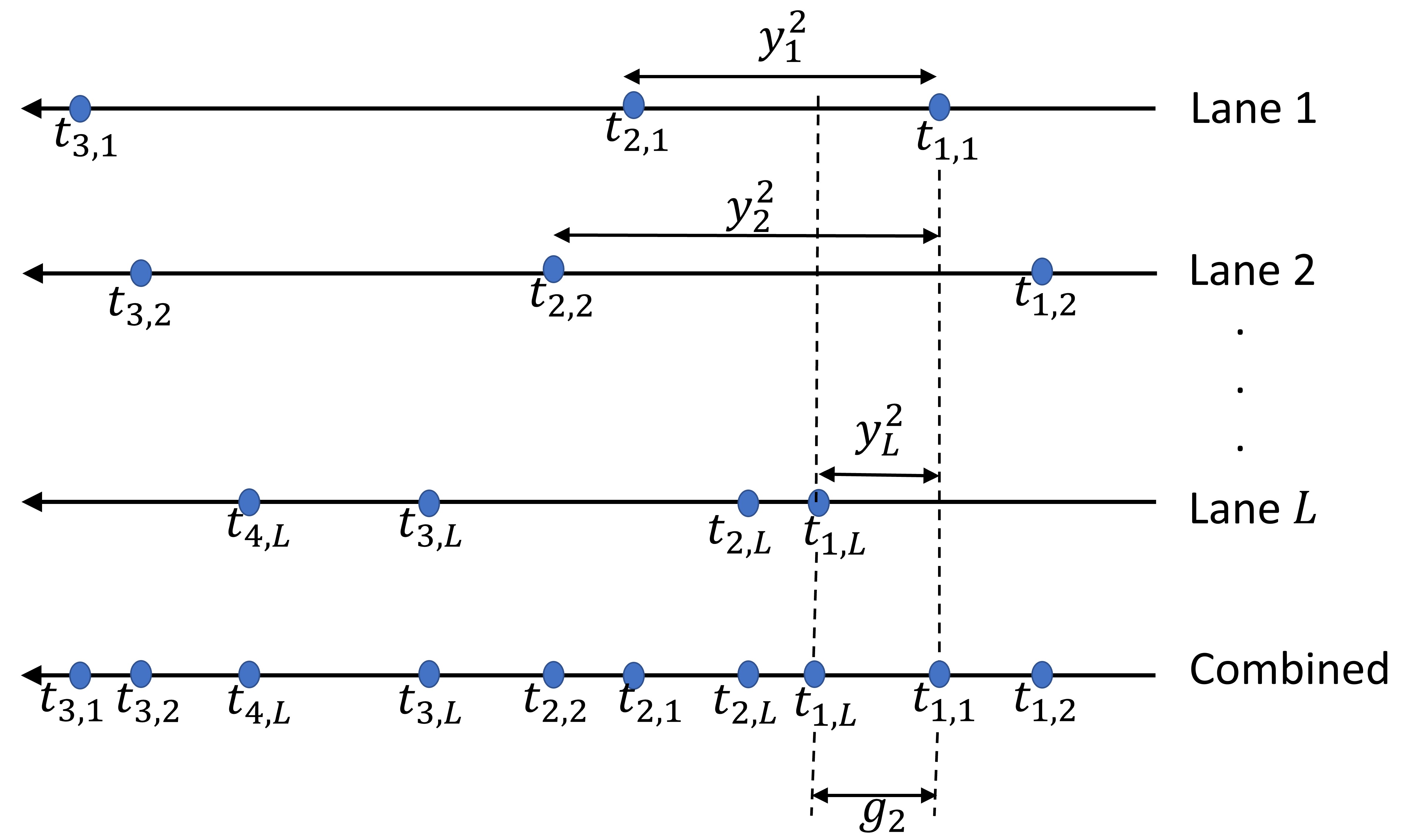}
  \caption{\label{g2}}
\end{subfigure}
\begin{subfigure}{\textwidth}
  \centering
  \includegraphics[width=0.6\textwidth]{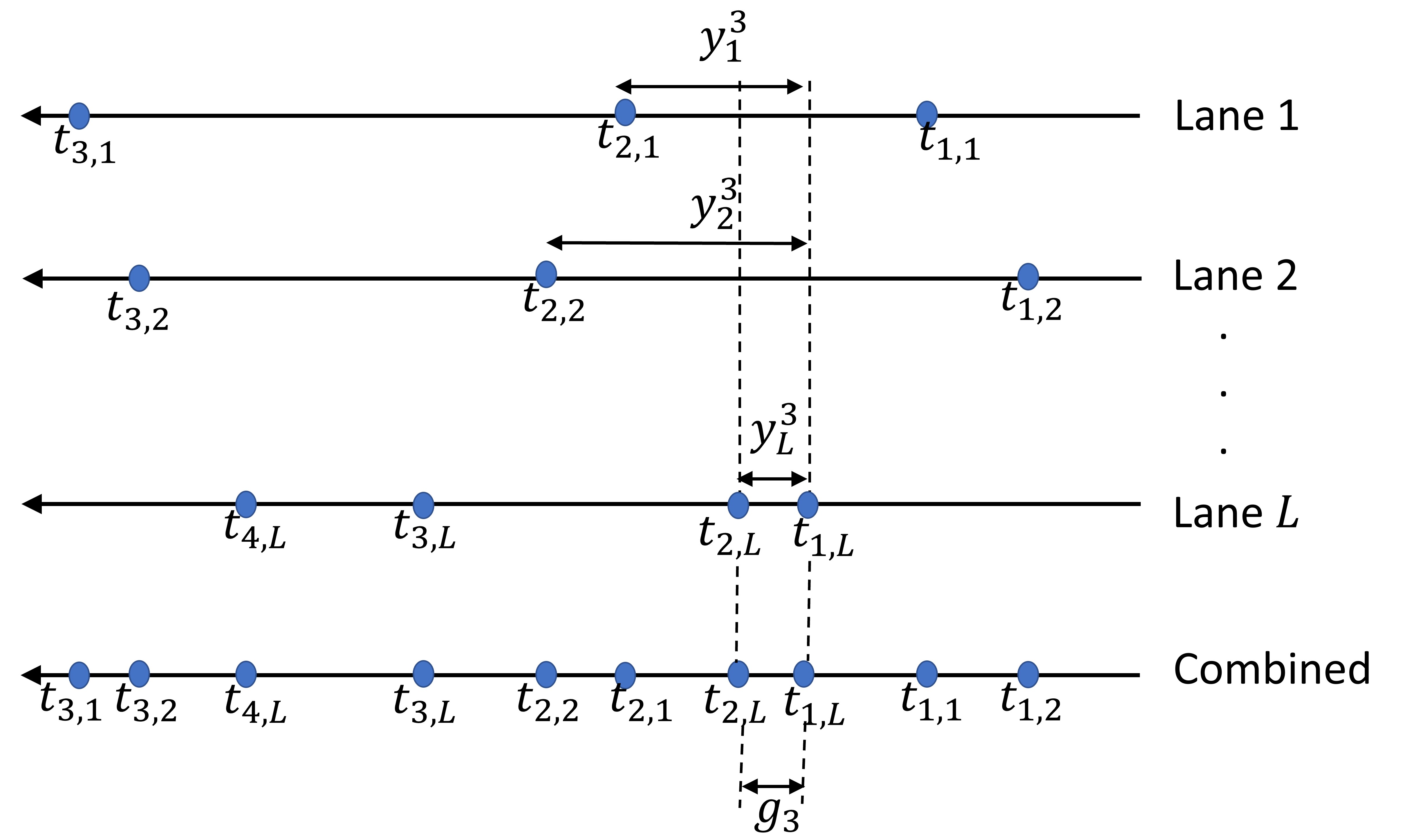}
  \caption{\label{g3}}
\end{subfigure}
\end{center}
\caption{\label{fig:all_gap} Representation of the gaps (a) first gap (gap 1), $g_1$, (b) second gap (gap 2), $g_2$ and (c) third gap (gap 3), $g_3$}
\end{figure}

To explain the new representation of gaps being proposed in this section Figure \ref{fig:all_gap} is introduced. In the figure, Figure \ref{fig:all_gap}(a), \ref{fig:all_gap}(b) and \ref{fig:all_gap}(c) each have four parallel lines representing time increasing from right to left. The first line shows the times at which successive vehicles of Lane 1 cross an arbitrarily located Section $\mathbb{S}$ (see Figure \ref{fig:gap_fig1}(b)) while the second line shows the same for Lane 2 and so on. The last line, however, shows when vehicles cross Section $\mathbb{S}$ irrespective of their lanes. This line is the same as the line drawn in \ref{fig:gap_fig1}(c). 

Let, $t^e$ be the time the first vehicle crosses Section $\mathbb{S}$ on the road. That is, 
\vspace{-12pt}
\[t^e = \min_{i,j}\ (t_{i,j}) \]
For the example shown in Figure \ref{fig:all_gap}, $t^e = t_{1,2}$.\\
Let, $y^1_j = (\text{earliest crossing time (after} \  t^e) \  \text{of vehicles in Lane}\ j)  - t^e$\\
For the example in Figure \ref{fig:all_gap}, $y^1_1,\ y^1_2, \cdots y^1_L$ are marked in Figure \ref{fig:all_gap}(a). The first gap $g_1$ can be obtained as: 
\vspace{-12pt}
\[ g_1 = \min (y^1_1,\ y^1_2, \cdots y^1_L) \] 
Note, this $g_1$ will be the same as the one obtained using the representation suggested in the previous section (see Figure \ref{fig:gap_fig1}(c)). \\
Let, $y^2_j = (\text{earliest crossing time (after} \  g_1+t^e)\  \text{of vehicles in Lane} j) - (g_1 + t^e) $\\
The quantities, $y^2_1,\ y^2_2, \cdots y^2_L$ are marked in Figure \ref{fig:all_gap}(b). The second gap $g_2$ can be obtained as: 
\vspace{-12pt}
\[ g_2 = \min (y^2_1,\ y^2_2, \cdots y^2_L) \]
As before, this $g_2$ will be the same as the one shown in Figure \ref{fig:gap_fig1}(c). Similarly,\\
let, $y^3_j = (\text{earliest crossing time (after} \  g_1 + g_2 + t^e) \  \text{of vehicles in Lane} j)  - (g_1 + g_2 + t^e) $\\
The quantities, $y^3_1,\ y^3_2, \cdots y^3_N$ are marked in Figure \ref{fig:all_gap}(c). The third gap $g_3$ can be obtained as: 
\vspace{-12pt}
\[ g_3 = \min (y^3_1,\ y^3_2, \cdots y^3_L) \] This $g_3$ will be same as the one in Figure \ref{fig:gap_fig1}(c).\\
Then, with the initialization of $g_0 = 0$, $y^k_j$ can be defined as follows:
\vspace{-12pt}
\[y^k_j = (\text{earliest crossing time (after} \  \sum_{s=0}^{k-1} g_s + t^e) \ \text{of vehicles in Lane} j)  - \left(\sum_{s=0}^{k-1} g_s + t^e \right) \]
Now, the $k^{\text{th}}$ gap $g_k$ for any $k$ can be obtained as: 
\vspace{-12pt}
\[ g_k = \min (y^k_1,\ y^k_2, \cdots y^k_L) \] 
Note, there is no distinction between the definition of $g_k$ presented here and that presented earlier because $\left(\sum_{s=0}^{k-1} g_s + t^e \right)$ is equal to the $k^{th}$ entry on the last line of Figure \ref{fig:all_gap}(a), \ref{fig:all_gap}(b) and \ref{fig:all_gap}(c) and time line of Figure \ref{fig:gap_fig1}(c). 

The reason for introducing, what might seem like, a convoluted definition for gaps is that it expresses gaps in terms of variables $y^k_j$ that are related to (but not necessarily equal to) headways in each lane. Now, in each lane, headways are assumed to be independently and identically distributed, therefore the lane-wise arrival process is a renewal process. The combined arrival process (like the one shown by the last line of Figure \ref{fig:all_gap}(a), \ref{fig:all_gap}(b) and \ref{fig:all_gap}(c)) is therefore a superposition of the renewal processes describing the lane-wise arrival. One can now rely on the theory of renewal processes (see, \citealt{cox_1962},\ \citealt{TaylKarl98}) to describe the gaps in the combined arrival process shown in Figure \ref{fig:all_gap}. 

\begin{figure}[h]
    \centering    
    \includegraphics[width=0.7\textwidth]{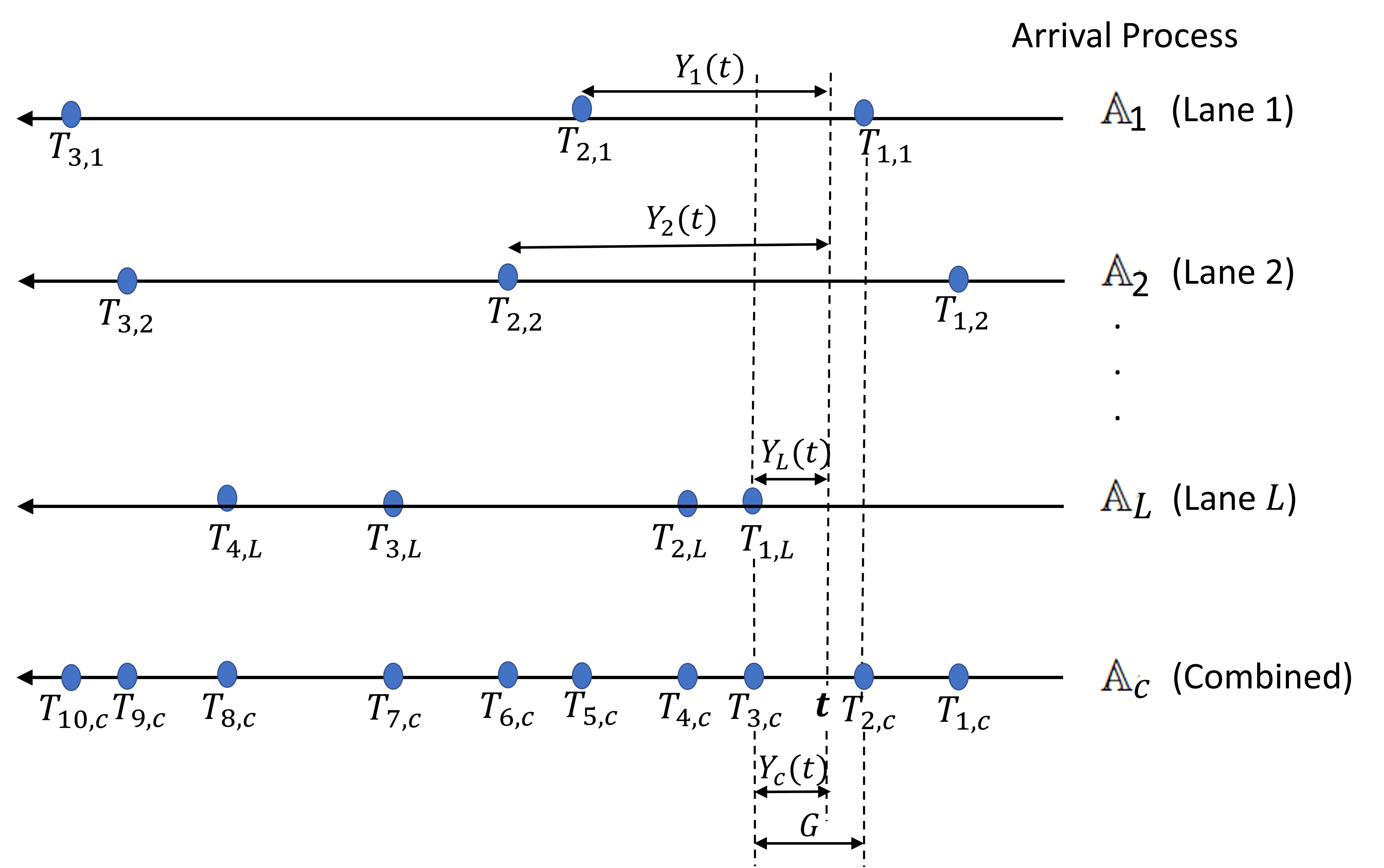}
    \caption{Gap $(G)$ and remaining time $Y_c(t)$ in combined arrival process}
    \label{fig:deriv}
\end{figure}

Although Figure \ref{fig:all_gap} poses the problem as an example instance, it is essentially a stochastic process since the lane-wise arrivals are all random. Figure \ref{fig:deriv} generalizes the representation; hence, in this representation, the arrival times in each lane are shown using upper case letters indicating that they are random quantities.  In the figure, the vehicle arrival process of Lane $j$ is denoted as $\mathbb{A}_j$ (where, $j \in (1,2,\cdots, L$); $L$ is the total number of lanes). Vehicle arrival process on the entire road, that is, the combined arrival process is denoted as $\mathbb{A}_c$. The combined arrival process is not a process that is separate from the processes $\mathbb{A}_1, \mathbb{A}_2, \cdots, \mathbb{A}_L$. It is a superposition of the arrival processes in the individual lanes. For example, in Figure \ref{fig:deriv}, $T_{1,c} = T_{1,2}, \ T_{2,c} = T_{1,1},\ T_{3,c} = T_{1,L}$ and so on. Here, $T_{i,j}$ is a random variable denoting the arrival time of the $i^{th}$ vehicle in $j^{th}$ lane (process). Before deriving the cdf and pdf of gaps a few more notation are introduced. 

Let, $N_j(t)$ be the number of vehicles that arrive till time $t$ in arrival process $\mathbb{A}_j$; that is, $N_j(t) = \sup \{i : T_{i,j} \leq t\}$. At time $t$, the "remaining time" on any arrival process $\mathbb{A}_j$, $Y_j(t)$, is defined as, 
\[
Y_j(t) = T_{N_j(t)+1,j}-t
\] 

Let, $F_{Y_j(t)}(y)$ and $f_{Y_j(t)}(y)$ be the cdf and pdf of remaining time (on the process $\mathbb{A}_j$) at time $t$. Since it can be shown that for large $t$, the distribution of $Y_j(t)$ is independent of $t$ (see, for example, \citealt{medhi1984stochastic}), the following simplified notation are respectively, used for the cdf and pdf: $F_{Y_j}(y)$, $f_{Y_j}(y)$\footnote{For the present problem of distribution of gaps, since the arrivals at any traffic location have been taking place over a long period time, $t$ can be assumed to be large.}. 

Let, $H_j$ be the inter-arrival time (headway) of process $\mathbb{A}_j$ and $F_{H_j}(h)$ and $f_{H_j}(h)$ be the cdf and pdf, respectively, of $H_j$; further let, $\mu_j = E[H_j]$. Recall, $\mathbb{A}_j$ is simply the arrival process of vehicles in a lane. From an understanding of renewal processes (\citealt{TaylKarl98}), the cdf, $F_{Y_j}(y)$, and pdf, $f_{Y_j}(y)$, of remaining time can be written as:
\begin{equation} \label{eq:cdf_rem_head}
         F_{Y_j}(y) = \frac{1}{\mu_j} \int_{0}^y [1-F_{H_j}(h)] dh
\end{equation}
\begin{equation} \label{eq:pdf_rem} 
         f_{Y_j}(y) = \frac{1}{\mu_j} [1-F_{H_j}(y)]
\end{equation}
As shown earlier and in Figure \ref{fig:deriv}, at any time $t$, the remaining time on the combined process is given by,
\[ Y_c(t) = \min(Y_1(t), Y_2(t), \cdots ,Y_L(t)) \]
Hence,
\begin{equation*}
    \begin{split}
     P(Y_c(t) \le y)
           &= P(\text{at least one of $Y_1(t), Y_2(t), \cdots, Y_L(t)$ } \le y)\\
           &= 1 - P(Y_1(t)>y, Y_2(t)>y,\cdots, Y_L(t)>y)
    \end{split}
\end{equation*}
Therefore, for large $t$, the cdf of remaining time on the combined process can be written as,
\begin{equation}
    F_{Y_c}(y) = 1-[(1-F_{Y_1}(y))(1-F_{Y_2}(y))\cdots(1-F_{Y_L}(y))]
\end{equation}
and the pdf can be written as,
\begin{align}
    f_{Y_c}(y) &= \frac{d}{dy} F_{Y_c}(y) = -\left[\sum_{j=1}^L \left(-f_{Y_j}(y)\right) \left(\prod_{k = 1, k\neq j}^L (1-F_{Y_k}(y))\right)\right]  \nonumber
\end{align}
Using Equation \ref{eq:pdf_rem}, $f_{Y_j}(y)$ can be replaced and the following can be written,
\begin{equation} \label{eq:pdf_rem_gap}
    f_{Y_c}(y) = \sum_{j=1}^L \frac{1}{\mu_j} \left(1-F_{H_j}(y)\right) \left(\prod_{k = 1, k\neq j}^L (1-F_{Y_k}(y))\right)
\end{equation}
Let, $F_G(g)$ be the cdf of gaps (that is the inter-event time for the combined process) and $\mu_g$ be the expected value of gaps. By rearranging the terms of Equation \ref{eq:pdf_rem} and noting that the flow on the road (combined process) is equal to the sum of the flows in the individual lanes, $F_G(g)$ can be written as,

\begin{align*}      
    F_{G}(g) &= 1-\mu_g f_{Y_c}(g) = 1-\frac{1}{\sum_{s=1}^L \frac{1}{\mu_s}} f_{Y_c}(g) \nonumber\\
             &= 1- \sum_{j=1}^L \left(1-F_{H_j}(g)\right) \left(\prod_{k = 1, k\neq j}^L \left(1-F_{Y_k}(g)\right)\right) \left(\frac{\frac{1}{\mu_j}}{\sum_{s=1}^L \frac{1}{\mu_s}}\right)  &\text{(from Equation \ref{eq:pdf_rem_gap})} \nonumber
\end{align*}
Using Equation \ref{eq:cdf_rem_head}, the above expression can be written as,
\begin{align} \label{eq:cdf_gap}    
       F_{G}(g)      &= 1- \sum_{j=1}^L \left(1-F_{H_j}(g)\right) \left(\prod_{k = 1, k\neq j}^L \left(1-\frac{1}{\mu_k} \int_{0}^g \left(1-F_{H_k}(h)\right) dh \right)\right) \left(\frac{\frac{1}{\mu_j}}{\sum_{s=1}^L \frac{1}{\mu_s}}\right) 
\end{align}

Therefore, the pdf of gaps, $f_G(g)$, can be written as,
\begin{align} \label{eq:pdf_gap}
    f_G(g) &= \frac{d}{dg} F_{G}(g) \nonumber\\
           &= \sum_{j=1}^{L} \Bigg[  \left(\frac{\frac{1}{\mu_j}}{\sum_{s=1}^{L}\frac{1}{\mu_s}}\right) \Bigg(f_{H_j}(g) \left(\prod_{k = 1, k\neq j}^{L} (1-F_{Y_k}(g))\right) +(1-F_{H_j}(g)) \nonumber\\
           & \quad \quad \quad \quad \quad \quad \quad \quad \quad \quad  \left(\sum_{k=1,k\neq j}^L (f_{Y_k}(g)) \prod_{m = 1,m \neq j,k}^L (1-F_{Y_m}(g))\right)  \Bigg)  \Bigg]
\end{align}

The development of Equation \ref{eq:pdf_gap} tacitly assumes that the movement in the stream is orderly and is divided into a known number of lanes (that is, $L$ is known). Next, the derivation is extended to include disorderly streams. 

While developing Equation \ref{eq:pdf_gap} it helped to think of vehicles of a particular arrival process (renewal process) to be physically separated from other such processes going on concurrently. These separations were referred to as lanes. A little reflection on the derivation process of Equation \ref{eq:pdf_gap} indicates that what is important are the arrival times and not the lateral locations. The spatial dimension plays no role. Thus, in this abstract world of renewal processes in time and their superposition, orderliness in space is not a requirement. One can imagine, as is done here, that a disorderly stream also consists of multiple renewal processes whose superposition gives rise to the gaps. Therefore, the pdf of gaps in such streams is also given by Equation \ref{eq:pdf_gap}. However, in this abstraction, unlike in the case of orderly streams, $L$ is an unknown parameter and will have to be estimated from data on gaps of such disorderly streams. 

A little reflection shows that $L$ represents the number of renewal processes required to best recreate the observed gaps. Thinking of $L$ as the number of lanes of an orderly stream, while making the process easily comprehendible, is a rather restrictive view. So, even for orderly streams, if one were to assume $L$ to be unknown and chose to estimate it from the data the results may yield a value of $L$ that is different from the number of lanes and such a result should not be surprising. 

Before leaving this section a few properties of renewal processes and their superposition relevant to this study need to be mentioned.

The theory of renewal processes suggests that as $L \rightarrow \infty$ (or when $L$ is large) then the superposed process (like $\mathbb{A}_c$ here) becomes a Poisson process (\citealt{khintchine}, \citealt{cinlar1972}). One may argue that for disorderly streams, for anything more than light flow, $L$ should be large (thinking of a disorderly stream as one that consists of a series of virtual lanes may help visualize this argument better). If this is true then one should observe that the gaps in disorderly flows are always exponentially distributed. It remains to be seen whether empirical observations support this line of argument. 

Next, attention is drawn to the fact that if the individual renewal processes (like $\mathbb{A}_1,\ \mathbb{A}_2,\ \cdots,\ \mathbb{A}_L$ here) are Poisson then the superposed process is also Poisson. This would imply if $F_{H_1}(h)$, $F_{H_2}(h)$, $\cdots$, $F_{H_L}(h)$ are exponential then $F_G(g)$ is also exponential. A quick extension of Equation \ref{eq:cdf_gap} will show that if $F_{H_j}(h) = 1 - e^{-\lambda_jh}$ then 
\vspace{-24pt}
\begin{equation} \label{eq:cdf_gap(exp)}
    F_G(g) = 1 - e^{- \left(\sum\limits_{j=1}^L \lambda_j\right) g}
\end{equation}
So, one can use Equation \ref{eq:cdf_gap(exp)} instead of Equation \ref{eq:cdf_gap} to represent the distribution of gaps if one assumes that the individual lanes have Poisson arrivals.

Finally, note that writing the likelihood as the product of probability terms in Equation \ref{eq:pdf_gap} assumes that the superposed point process on which gaps are defined is a renewal process. Although this is true when $L$ is large or individual renewal processes that generate the superposed point process are Poisson; in general one cannot claim that the superposed process is a renewal process. Ideally one needs to check whether, in their case, the superposed process can be assumed to be a renewal process. The test suggested by \citet{cox_lewis_1966} can be employed to do the check.

Before leaving this section, it is worth mentioning that the development of Equation \ref{eq:cdf_gap} presented here closely follows how the superposition of renewal processes are typically handled (see, \citealt{CoxandSmith}). A slightly different approach to deriving Equation \ref{eq:cdf_gap}, developed as a part of this work, is shown in Appendix \ref{app:distribution}.

\section{Description of the Data} \label{sec:data}
In order to investigate how effective Equation \ref{eq:pdf_gap} is as a representation of pdf of gaps, data from three traffic streams (one orderly stream and two disorderly streams) from three different locations are used in this study. For orderly streams, about 60 minutes of vehicle trajectory data from a three-lane mid-block section in  Cologne, Germany, available in open source HighD dataset (\citealt{highDdataset}), is used. A total of about 5000 vehicles are observed. The times ($t_{i,j}$) when vehicles cross a given section are obtained from the data. These are then used to find headways (in each lane) and calculate gaps as described in Section \ref{prob_stat}. Table \ref{tab:des_head} shows the descriptive statistics of headways in each lane and gaps for this site.
\begin{table}[h]
\begin{center}
  \caption{\label{tab:des_head} Description of headways and gaps in orderly stream}
  \begin{tabular}{ccK{2cm}K{2cm}}
  \hline
  \multirow{2}{*}{Lane} & Observed Number of Headways &  \multicolumn{2}{l}{Headway/Gaps (in seconds) }\\   \cline{3-4} 
      & (or Gaps for Combined Process) &  Mean & Std. Dev.\\
    \hline
    1 &1162  &2.72 &1.80\\
    2 & 1826 &1.73 &1.02\\
    3 & 2012 &1.57 &0.96\\
    Combined & 5000 &0.63 &0.51\\
    \hline
  \end{tabular}  
\end{center}
\end{table}
For disorderly streams, traffic data is collected from two Indian cities, namely, Kanpur and Chennai. 
The first data collection site is on a three-lane wide highway in Kanpur. The second data is from Chennai on a four-lane wide highway. (Note, for disorderly streams the number of lanes is mentioned only to indicate how wide the roads were.)

Gaps are determined as the difference between successive arrivals at a section; the lateral positions of vehicles are ignored. Table \ref{tab:gap_data} shows the details of gaps for both locations. The flow on the roads of Kanpur and Chennai are observed as 2117 veh/hr and 4264 veh/hr, respectively.

\begin{table}[h!]
\begin{center}
  \caption{\label{tab:gap_data} Description of gaps in disorderly streams}
  \begin{tabular}{ccK{2cm}K{2cm}}
  \hline
  \multirow{2}{*}{Location} & Observed Number &  \multicolumn{2}{c}{Gap (in seconds) }\\   \cline{3-4} 
      &  of Gaps &  Mean & Std. Dev.\\
    \hline
    Kanpur  & 6640  & 1.70 & 1.97\\
    Chennai & 10767 & 0.84 & 0.91\\
    \hline
  \end{tabular}  
\end{center}
\end{table}
The next section describes the parameter estimation process for the pdf of gaps from such real-world data.

\section{Parameter Estimation Process} \label{sec:est_process}
The parameters of the pdf can be estimated from gaps. Alternatively, for orderly streams where one can observe headways and where $L$ can be reasonably assumed, the parameters of pdf of gaps can be obtained from the estimated parameters of the headways. Both the approaches use MLE method. These approaches are described next.

Let the sample of observed gaps be $g_1,g_2,g_3,\cdots,g_{S}$, where $S$ is the number of gaps. The likelihood, $\mathcal{L}_g$, of observing the given sample of gaps is obtained using Equation \ref{eq:pdf_gap}. Mathematically it can be written as,\footnote{To check whether the superposed process (or gaps) studied here can be assumed to be renewal processes the test suggested by \citealt{cox_lewis_1966} is performed on multiple randomly chosen disjoint sets of gaps for each of the three sites. For almost all of these disjoint sets, the hypothesis that the gaps come from a renewal process could not be rejected at a 95 per cent level of significance.}
\begin{equation*}
    \begin{split}
    \mathcal{L}_g &= \prod_{s=1}^{S} f_G(g_s) \\
            &= \prod_{s=1}^{S} \Bigg[\sum_{j=1}^{L} \Bigg[  \left(\frac{\frac{1}{\mu_j}}{\sum_{s=1}^{L}\frac{1}{\mu_s}}\right) \Bigg(f_{H_j}(g_s) \left(\prod_{k = 1, k\neq j}^{L} (1-F_{Y_k}(g_s))\right) +(1-F_{H_j}(g_s)) \nonumber\\
            & \quad \quad \quad \quad \quad \quad \quad \quad \quad \quad  \Bigg(\sum_{k=1,k\neq j}^L (f_{Y_k}(g_s)) \prod_{m = 1,m \neq j,k}^L (1-F_{Y_m}(g_s))\Bigg)  \Bigg)  \Bigg] \Bigg]
    \end{split}
\end{equation*}
The log-likelihood, $\mathcal{LL}_g$, is given by:
\begin{align}\label{eq:LL_gap}
    \mathcal{LL}_g &= ln(\mathcal{L}_g) = \sum_{s=1}^{S} ln\left(f_G(g_s)) \right) \nonumber\\
       &= \sum_{s=1}^{S} ln \Bigg[\sum_{j=1}^{L} \Bigg[  \left(\frac{\frac{1}{\mu_j}}{\sum_{s=1}^{L}\frac{1}{\mu_s}}\right) \Bigg(f_{H_j}(g_s) \left(\prod_{k = 1, k\neq j}^{L} (1-F_{Y_k}(g_s))\right) +(1-F_{H_j}(g_s)) \nonumber\\
       & \quad \quad \quad \quad \quad \quad \quad \quad \quad \quad  \Bigg(\sum_{k=1,k\neq j}^L (f_{Y_k}(g_s)) \prod_{m = 1,m \neq j,k}^L (1-F_{Y_m}(g_s))\Bigg)  \Bigg)  \Bigg] \Bigg]
\end{align} 

Note that, in the case of a multi-lane orderly stream the parameters of the log-likelihood function are parameters of the assumed distribution of lane-wise headways (i.e., parameters of inter-event time in each renewal process). For example, if $H_j \sim \text{Gamma}(k_j,\lambda_j)$ then the parameters of the gap distribution are $k_j$ and $\lambda_j$ for $j \in (1,2,\cdots,L$). In the case of disorderly streams, additionally, the number of renewal processes ($L$) is also a parameter\footnote{More correctly one should say that the parameters are $L$ and the parameters of the $L$ constituent renewal processes required to recreate the superposed process or gaps distribution.}. As per the principle of MLE, the parameter values that give the maximum log-likelihood are the estimated parameter values of pdf of gaps.

The process described so far uses data on gaps to estimate the parameters of its pdf. In the case of disorderly streams, where gaps are the only observable quantities, this remains the only viable option to estimate the gap distribution parameters including $L$. In the case of orderly streams, however, the headways of the individual lanes (renewal processes) are also observable. Hence, if such data is available then, in theory, one could estimate the parameters of these distributions using MLE with log-likelihood given by $\mathcal{LL}_{h}$ in Equation \ref{eq:LL_head} that follows. The parameters estimated thus can be used in Equation \ref{eq:pdf_gap} to express $f_G(g)$. 

Let the observed headways in lane $j$ be $h_1,h_2,h_3,\cdots,h_{M}$, where $M$ is the number of headways in lane $j$. The log-likelihood function ($\mathcal{LL}_{h}$) of observing these headways in lane $j$ can be written as,
\begin{equation} \label{eq:LL_head}
    \mathcal{LL}_{h} = \sum_{m=1}^{M} ln\left(f_{H_i}(h_m)\right)
\end{equation}
As is apparent, this is an indirect method. The parameters of $f_G(g)$ are not obtained from gaps data but are reconstructed from parameter estimates of individual headway data. At best such a reconstruction can provide a reasonable approximation of the parameters. Hence, one should not expect the parameters obtained here to outperform the parameters estimated directly from the gaps data. 

Before leaving this section, a discussion on the estimability of $L$ is required. Note, $L$ (in Equation \ref{eq:LL_gap}) determines the number of renewal processes that need to be superposed. This implies that $L$ determines how many variables are required to express $f_G(g)$. In a way, $L$ specifies the exact function for $f_G(g)$. Since MLE or any other estimation process assumes that the function whose parameters are to be estimated is fully specified, a value of $L$ needs to be assumed to proceed with the estimation process. Since if $L$ changes the number of $k_j$ and $\lambda_j$ parameters change the maximum likelihood values obtained for different values of $L$ cannot be directly compared. Once the estimations for different values of $L$ are carried out, Akaike’s information criterion (AIC) are compared (for more details about AIC, see \citealt{aic}). The estimated parameter values for the $L$ that gives the minimum AIC are taken as the final estimated values. 

Lastly, it may be pointed out that sometimes it may also happen that MLE is unable to estimate all the relevant parameters because of the way they appear in the likelihood function. For example, if $L$ is the number of constituent renewal processes where the inter-arrival times for each process are exponentially distributed with parameter $\lambda_j$ thus the superposed process is also exponentially distributed with parameter $\sum \lambda_j$. As can be seen from Equation \ref{eq:cdf_gap}, in this case, MLE can estimate only $\sum \lambda_j$ and not the individual $\lambda_j's$.

\section{Estimates of Parameters and Discussions}
Parameter estimates of $f_G(g)$ shown in Equation \ref{eq:pdf_gap} using (i) the likelihood function described in the previous section and (ii) the data introduced in Section \ref{sec:data} are presented. The log-likelihood function is maximised using the maxLik package (\citealt{MaxLik}) available in R. The results presented in this section assume that each constituent renewal process has inter-arrival times (or headways in the case of orderly streams) that are distributed according to a Gamma distribution. That is, $F_{H_j}(h)$ is assumed to be a Gamma distribution with shape parameter $k_j$ and rate parameter $\lambda_j$. The reason for choosing Gamma distribution is that it is among a general class of distribution used for headway distribution (\citealt{Li_chen}). As in previous sections here also orderly streams are discussed first followed by disorderly streams.

\subsection{Parameter Estimates for Orderly Stream}
In orderly streams both gap data and individual lane headway data are available. So, both the methods for parameter estimation discussed in the previous section are used. 

\subsubsection{Estimating (or approximating) pdf of gaps from parameters estimated using headway data}
\begin{figure}[h]
    \centering
    \includegraphics[width=0.7\textwidth]{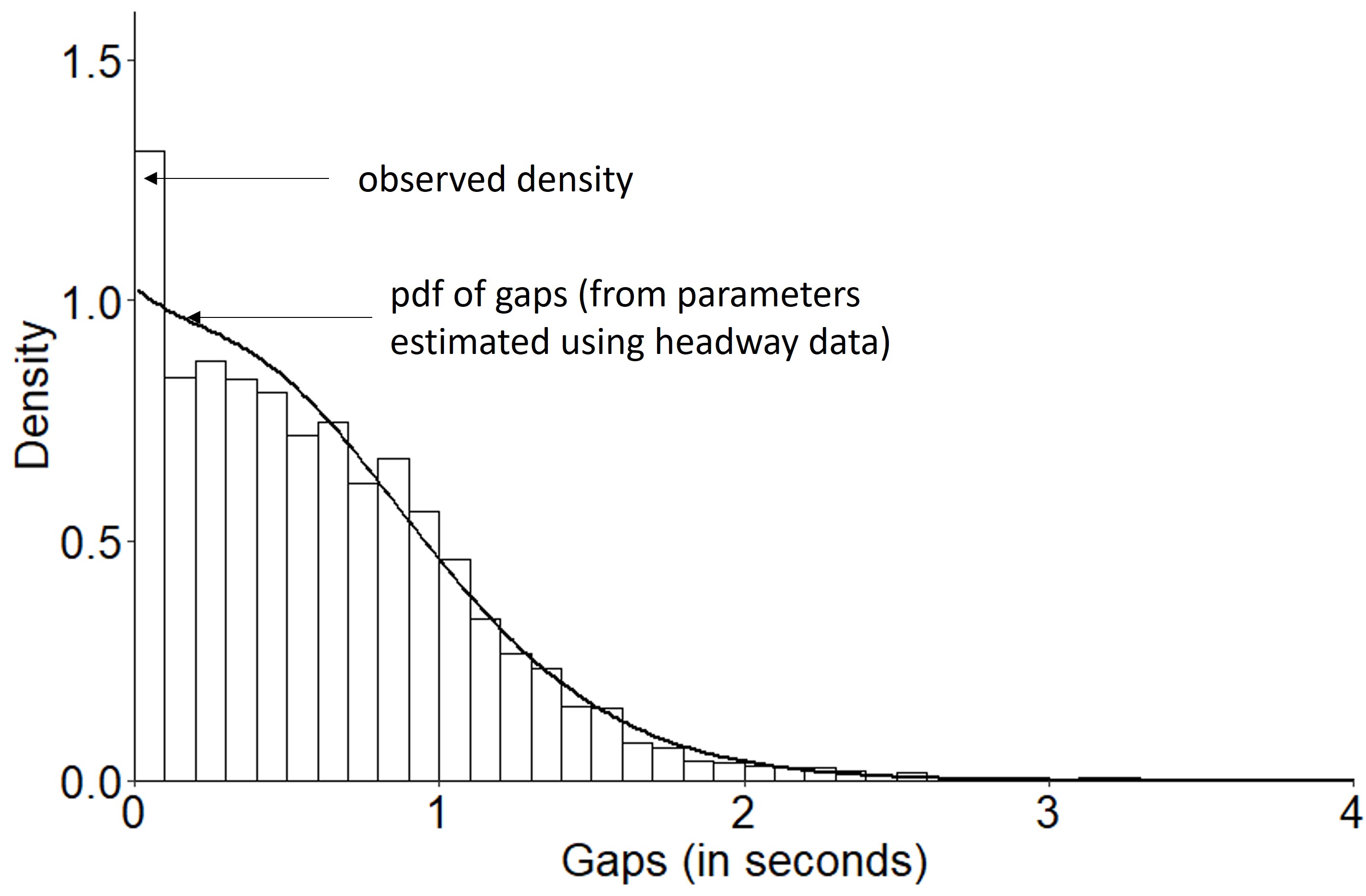}
    \caption{\label{fig:gap_gam_lane} Density histogram of observed gaps and plot of pdf of gaps for orderly streams using parameters estimated from headway data assuming Gamma distribution for headways.} 
\end{figure}

Table \ref{tab:gam_lane} presents the estimates obtained from maximising the log-likelihood function given by Equation \ref{eq:cdf_gap} for the data introduced in Table \ref{tab:des_head}. These $\hat{k_j}$ and  $\hat{\lambda_j}$ are used as a parameter estimates of $F_{H_j}(\cdot), f_{H_j}(\cdot), F_{Y_j}(\cdot)$ and $f_{Y_j}(\cdot)$ in $f_G(g)$ given by Equation \ref{eq:pdf_gap}. The observed density histogram of gaps and the plot of the pdf using estimated parameters is shown in Figure \ref{fig:gap_gam_lane}.

\begin{table}[h]
    \centering
    \caption{Estimated parameters from headway data assuming Gamma distribution for headways for orderly streams with three lanes (see Table \ref{tab:des_head})}
    \begin{tabular}{K{1 cm} c c}
    \hline
        $j$ & $\hat{k_j}$ (t-value)& $\hat{\lambda_j}$ (t-value) \\
        \hline
        1    & 2.741 (25.461) & 1.005 (23.203)  \\
        2    & 3.372 (31.624) & 1.945 (29.329) \\
        3    & 4.096 (32.944) & 2.604 (30.965) \\
        \hline
    \end{tabular}    
    \label{tab:gam_lane}
\end{table}

It can be seen from Figure \ref{fig:gap_gam_lane} that the gap distribution obtained using estimated headway parameters is possibly not the best; the line deviates from observations for small values of gaps. This can be due to two reasons. 

\begin{figure}[h]
    \centering
    \includegraphics[width=0.7\textwidth]{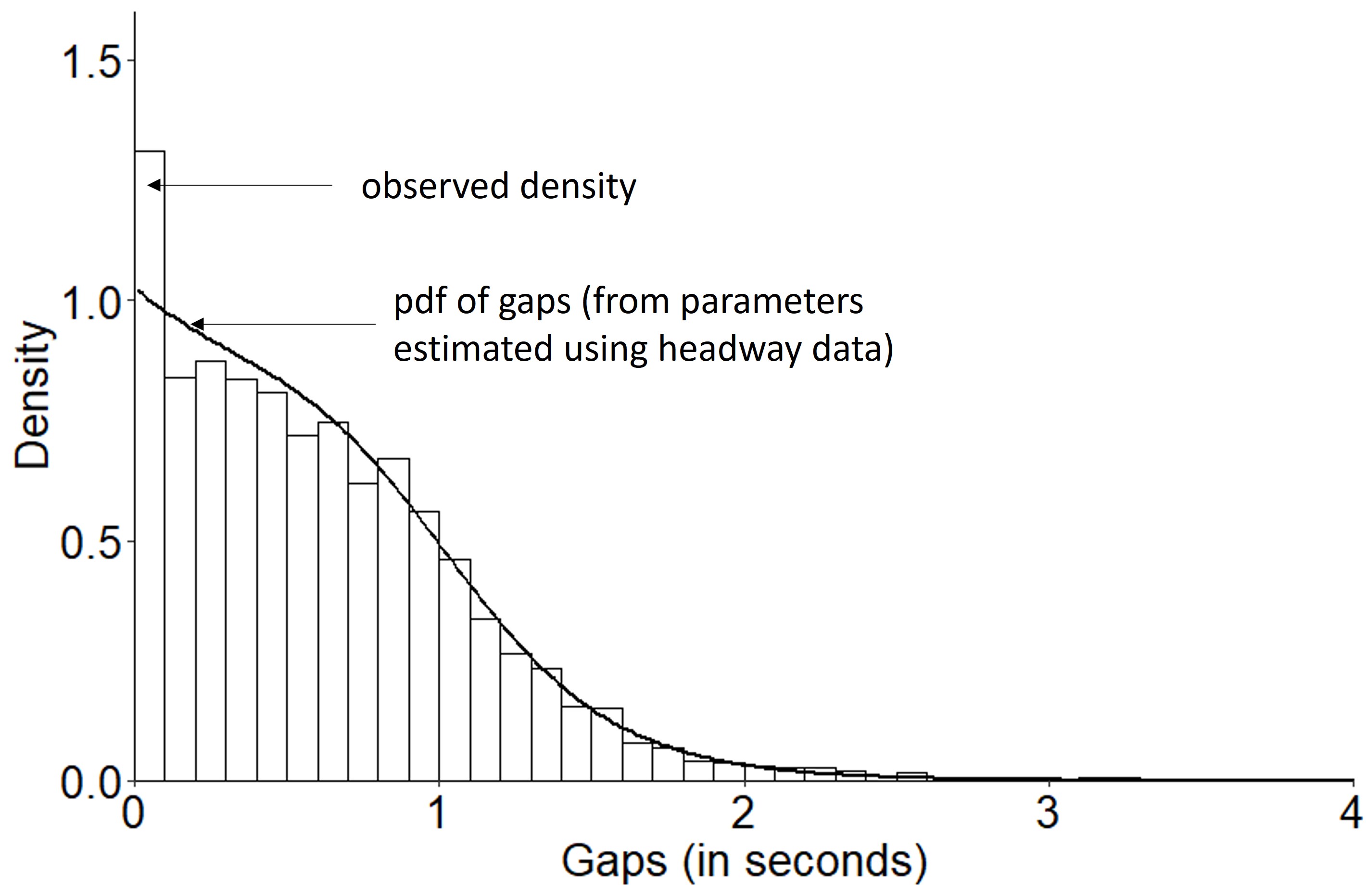}
    \caption{\label{fig:gap_llogis_lane} Density histogram of observed gaps and plot of pdf of gaps for orderly streams using parameters estimated from headway data assuming log-logistic distribution for headways} 
\end{figure}

First, while finding gaps distribution, one is assuming some distribution for the headways of each lane. Therefore, the final gap distribution is dependent on the assumed distribution of headway. One may need to fit different headway distributions to see whether the results improve. For example, for the data presented here, instead of Gamma distribution, log-logistic distribution for headway was also tried. Figure \ref{fig:gap_llogis_lane} shows the density histogram of observed gaps and the pdf using log-logistic distribution for headways. This pdf also suffers from the same shortcomings as the one shown in Figure \ref{fig:gap_gam_lane}. Some other attempts with other distributions also did not yield an estimated $f_G(g)$ that can be considered as a satisfactory representation of the data.

Second, and a more plausible reason for estimating an unsatisfactory pdf may be that the parameters used to obtain the pdf of gaps were not estimated from the gaps data but from headway data. Consequently, the $f_G(g)$ shown in Figure \ref{fig:gap_gam_lane} and Figure \ref{fig:gap_llogis_lane} are really not "estimated" but more like "approximated" using estimated parameters of headways. The pdf estimated using gaps data may perform better.

\subsubsection{Estimating pdf of gaps using gaps data} \label{sec:est_from_gap}
In this section, parameters of $f_G(g)$ are estimated directly from gaps data by maximising log-likelihood given in Equation \ref{eq:LL_gap}. 

Before presenting the estimated parameter values, attention is drawn to the discussion on $L$ given near the end of Section \ref{sec:dist_gap}. Recall, mathematically, $L$ is the number of constituent renewal processes that gives rise to the gap distribution (the superposed process). There is no reason to believe that for a given superposed process (gap distribution) there will be only one set of constituent renewal processes (with their unique value of $L$, $\lambda_j's$ and $k_j's$) that give rise to it\footnote{Appendix \ref{app:non_unique} gives an example to illustrate this.}. One may also see (in orderly streams) that the MLE process gives estimates of $L$ (and $\lambda_j's$ and $k_j's$) that match or predict the observed gap distribution well but $L$ different from the number of lanes.

\begin{table}[h!]
    \centering
    \captionof{table}{\label{tab:gap_order_L3} Estimated parameters from gaps data for the orderly stream with three lanes (see Table \ref{tab:des_head})}
    \begin{tabular}{K{0.8 cm} K{1cm} c c c}
    \hline
        \multirow{2}{*}{$L$} & \multirow{2}{*}{$j$} &\multirow{2}{*}{$\hat{k_j}$ (t-value)}  & \multirow{2}{*}{$\hat{\lambda_j}$ (t-value)} & Maximum \\
                         & &                 &               & log-likelihood \\
        \hline 
        \multirow{3}{*}{3}&1& 3.282 (16.129)   & 3.343 (22.503)       &\multirow{3}{*}{-2473.65}\\                    
                          &2& 0.501 (13.604)   & 0.280 (5.010) & \\
                          &3& 1.238 (NaN)      &3.410e-6(2.948e-5) & \\
        \hline
        \multirow{2}{*}{2}&1& 3.282 (15.338)    & 3.343 (19.437) & \multirow{2}{*}{-2473.65}\\
                          &2&0.501 (19.671)      &0.280 (10.868)  & \\
    \hline     
    \end{tabular}
\end{table} 

\begin{figure}[h]
    \centering
    \includegraphics[width=0.7\textwidth]{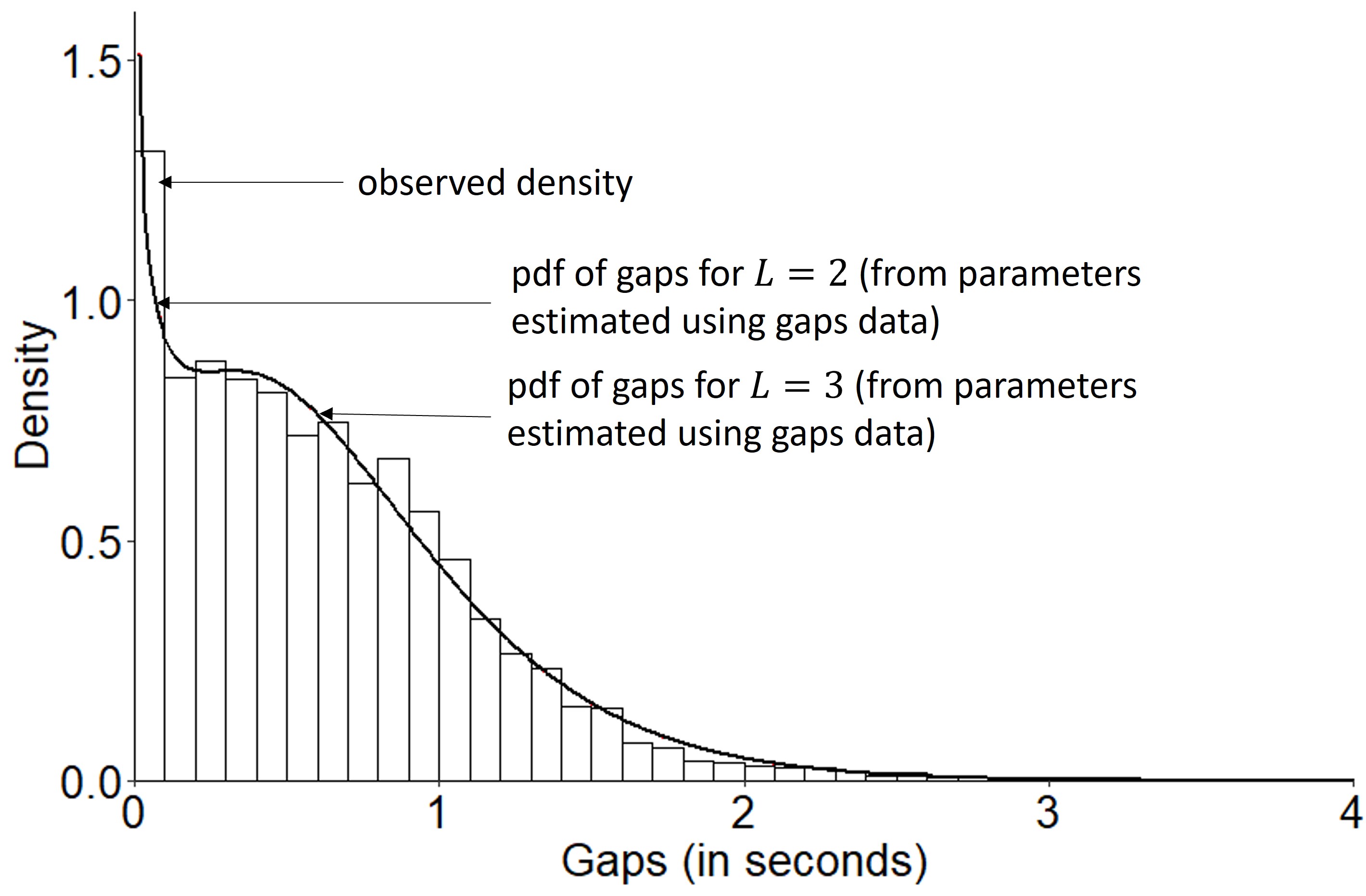}
    \caption{\label{fig:gap_order_L2} Density Histogram of observed gaps and plot of pdf of gaps for orderly streams (parameter estimation from gaps data)}
\end{figure}

The estimated parameters using Equation \ref{eq:LL_gap} and the data on gaps are shown in Table \ref{tab:gap_order_L3}. The table shows the parameter estimates with their t-values in parentheses for two different assumed values of $L$. Figure \ref{fig:gap_order_L2} presents a density histogram of observed gaps superimposed with $f_G(g)$ plotted using the estimated values of $\lambda_j$ and $k_j$ shown in Table \ref{tab:gap_order_L3}. Only one line is seen since the $f_G(g)$ for $L=3$ and $L=2$ overlap one another. 

The reason why only one line is seen is that there is no difference in the superposed process (gap distribution) obtained from either set of estimated parameters. The maximum log-likelihood value also indicates that both sets are indistinguishable (at least from the likelihood perspective). The value of $\hat{\lambda_3}$ (which is almost zero) also suggests one does not need three constituent processes to generate this gap distribution. Hence, it is not a surprise that the values of $\hat{\lambda_j}$ and $\hat{k_j}$ (for $j=1$ and $j=2$ for the $L=3$ case) are the same as the $\hat{\lambda_j}$ and $\hat{k_j}$ (for $j=1$ and $j=2$ for the $L=2$ case). The results reinforce the notion (put forward earlier) that, mathematically, $L$ is just the number of renewal processes required to be superposed to recreate the gaps distribution and it may not be the same as the number of lanes. 

Finally, it can be said that, not surprisingly (see earlier discussion), direct estimation of the $f_G(g)$ parameters from the gaps data yields a better fit than when the gap distribution is obtained from the parameters of the individual headway distributions. 

\subsection{Parameter Estimation for Disorderly Stream}
\begin{figure}[h!]
\begin{center} 
\begin{subfigure}{\textwidth}
  \centering
  \includegraphics[width=0.7\textwidth]{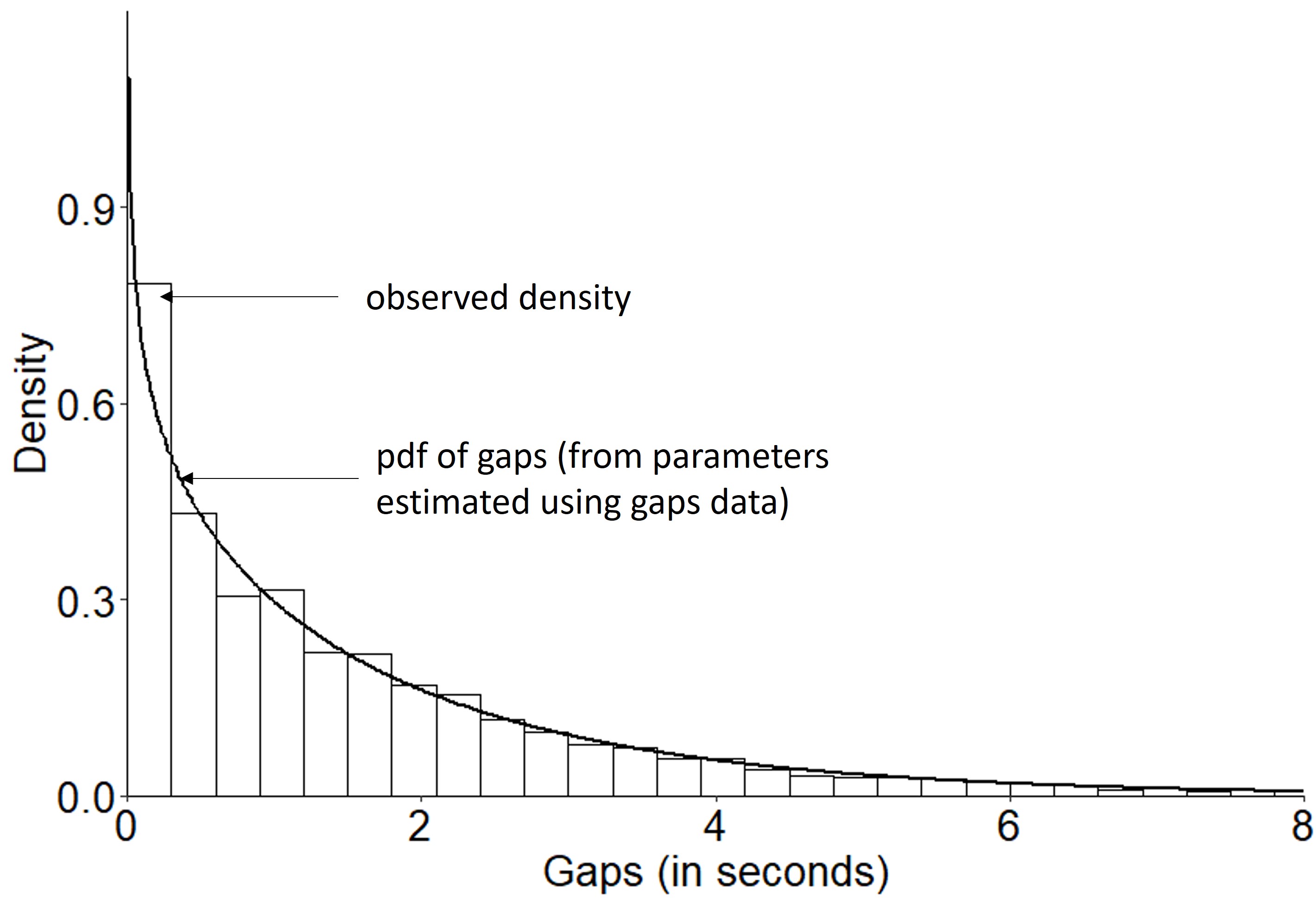}
  \caption{}
\end{subfigure}
\begin{subfigure}{\textwidth}
  \centering
  \includegraphics[width=0.7\textwidth]{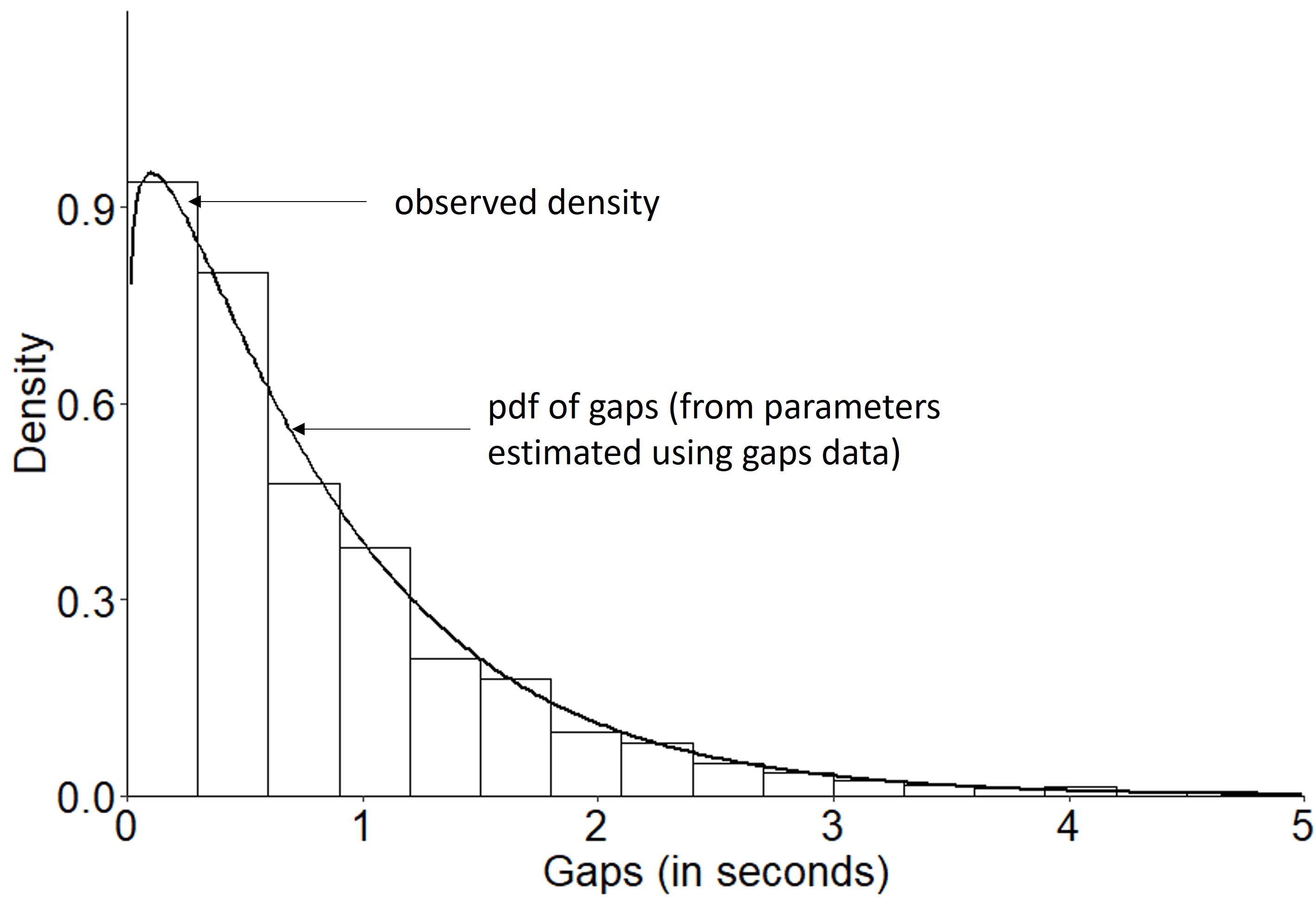}
  \caption{}
\end{subfigure}
\end{center}
\caption{\label{fig:disorder_gap} Density Histogram of observed gaps and plot of pdf of gaps for disorderly streams (parameter estimation from gaps data); (a) for Kanpur, (b) for Chennai.}
\end{figure}

In disorderly streams, there is no concept of headway. So, parameters of $f_G(g)$ can be estimated from gaps data only. As discussed in Section \ref{sec:dist_gap}, here in addition to $k_j$ and $\lambda_j$ ($j \in 1,2,\cdots,L$), $L$ also has to be estimated. Hence, the log-likelihood is maximised for integer values of $L$ from 1 to 5 for both Kanpur and Chennai data. For a given dataset, the AIC value for each $L$ is compared and the parameter estimates for the $L$ that gives the minimum AIC value are chosen as the final estimates. These estimated parameters are presented in Table \ref{tab:gap_disorder} and corresponding density plots are shown in Figure \ref{fig:disorder_gap}. The $k_j$ and $\lambda_j$ estimates for other values of $L$ together with their maximum log-likelihood values are provided in Appendix \ref{app:table}.

\begin{table}[h]
    \centering
    \captionof{table}{\label{tab:gap_disorder} Estimated parameters from gaps data for disorderly streams (see Table \ref{tab:gap_data})}
    \begin{tabular}{c c c c}
    \hline
        Location & $\hat{L}$ & $\hat{k}_1$ (t-value) & $\hat{\lambda}_1$ (t-value) \\
        \hline 
        Kanpur  & 1  &  0.820 (66.634) &  0.482 (49.453)\\
        Chennai & 1  &  1.169 (82.157) &  1.414 (66.258)\\
        \hline
    \end{tabular}
\end{table}

Note that, $L=1$ implies that gaps are gamma distributed. Further, for either of the data sets (see Table \ref{tab:gap_disorder}) the estimated $\hat{k}$ value is near unity. That is, the estimated Gamma distribution for the gap can be well approximated by an exponential distribution (recall, a Gamma distribution with $k=1$ is an exponential distribution). This observation lends support to the discussion in Section \ref{sec:dist_gap} where it was argued that gap distribution in disorderly streams should tend towards an exponential distribution. 

\section{Conclusion}
Gap acceptance is a fundamental aspect of how vehicles of contiguous or opposing traffic streams interact. Since the 1950s, researchers have paid substantial attention to mathematically modelling gap acceptance behaviour (for example, \citealt{tanner1951}, \citealt{daganzo1977} and others). Strangely, not enough attention was paid to describing the distribution of gaps. More attention was paid to understanding how headways are distributed. Of course, the fact that headways are not the same as gaps for multi-lane orderly streams was realised in the past too (see, \citealt{tanner1951}, \citealt{TIAN1999187}). Researchers simply assumed a distribution for gaps and proceeded with the analysis of how gap acceptance works in such an environment. For disorderly streams, the lack of understanding of the distribution of gaps is even more serious since in such streams the idea of headway is not meaningful. Yet, as in the case of orderly streams, here also, researchers just assume a distribution for gaps. In neither case was an attempt made to put forward a basis for the choice of the distribution of gaps one assumed. Hence, from the standpoint of gap acceptance, the importance of studying gap distribution cannot be overemphasized.

With the advent of powerful computers, there is an increasing reliance on simulation as a tool for understanding traffic movements. While one can generate incoming traffic for orderly streams from lane-based headway distributions, the same is not possible for disorderly streams since the notion of headways does not exist for such streams. In these latter types of streams, traffic can only be generated using gap distributions. (Surprisingly, most of the simulation studies on disorderly streams do not even discuss how the traffic is generated and the few that do simply state that they have used some headway distribution.) Thus, a study on gap distributions is important for developing reasonable simulations of disorderly streams.

In this work, an attempt has been made to analytically determine the distribution of gaps under certain assumptions on the vehicle arrival process. The theory of renewal processes is exploited to develop an expression for the distribution of gaps by realising that gaps are inter-event times in a superposed renewal process. Parameters of the gap distribution obtained through such a process can be estimated using an MLE framework. Results from fitting this distribution to real-world data on gaps from orderly streams show that it can mimic the peculiar shape of the distribution that one observes (see Figure \ref{fig:gap_order_L2}). Real-world data from various locations for disorderly streams indicate that gaps seem to be exponentially distributed. Interestingly, the theoretical development of gap distribution for such streams proposed here also indicates that gaps for disorderly streams on wide roads should have an exponential distribution.

The expression for the distribution of gaps for multi-lane orderly and disorderly streams developed here can only be evaluated numerically. This also poses challenges during parameter estimation. Although this remains a non-ideal situation, with today's computational power its use in practical problems will not be restricted. 

In short, the contributions of this work are: (i) developing a framework to analytically derive the distribution of gaps for multi-lane orderly and disorderly streams, (ii) providing a basis for specifying a gap distribution, either analytically or numerically, for gap acceptance studies in orderly and disorderly streams and (iii) strengthening the conclusions drawn from simulation studies of disorderly streams by facilitating the development of a theoretically sound vehicle arrival module necessary in such simulations.

\section*{Acknowledgements}
The Chennai dataset used in this study was made available by Prof. Venkatesan Kanagaraj of Indian Institute of Technology Kanpur.

\bibliography{ref}

\begin{appendix}
\section{Determination of pdf of Gaps: Alternate Approach to Deriving Equation {\ref{eq:cdf_gap}}} \label{app:distribution}
\begin{figure}[H]
    \centering    
    \includegraphics[width=0.7\textwidth]{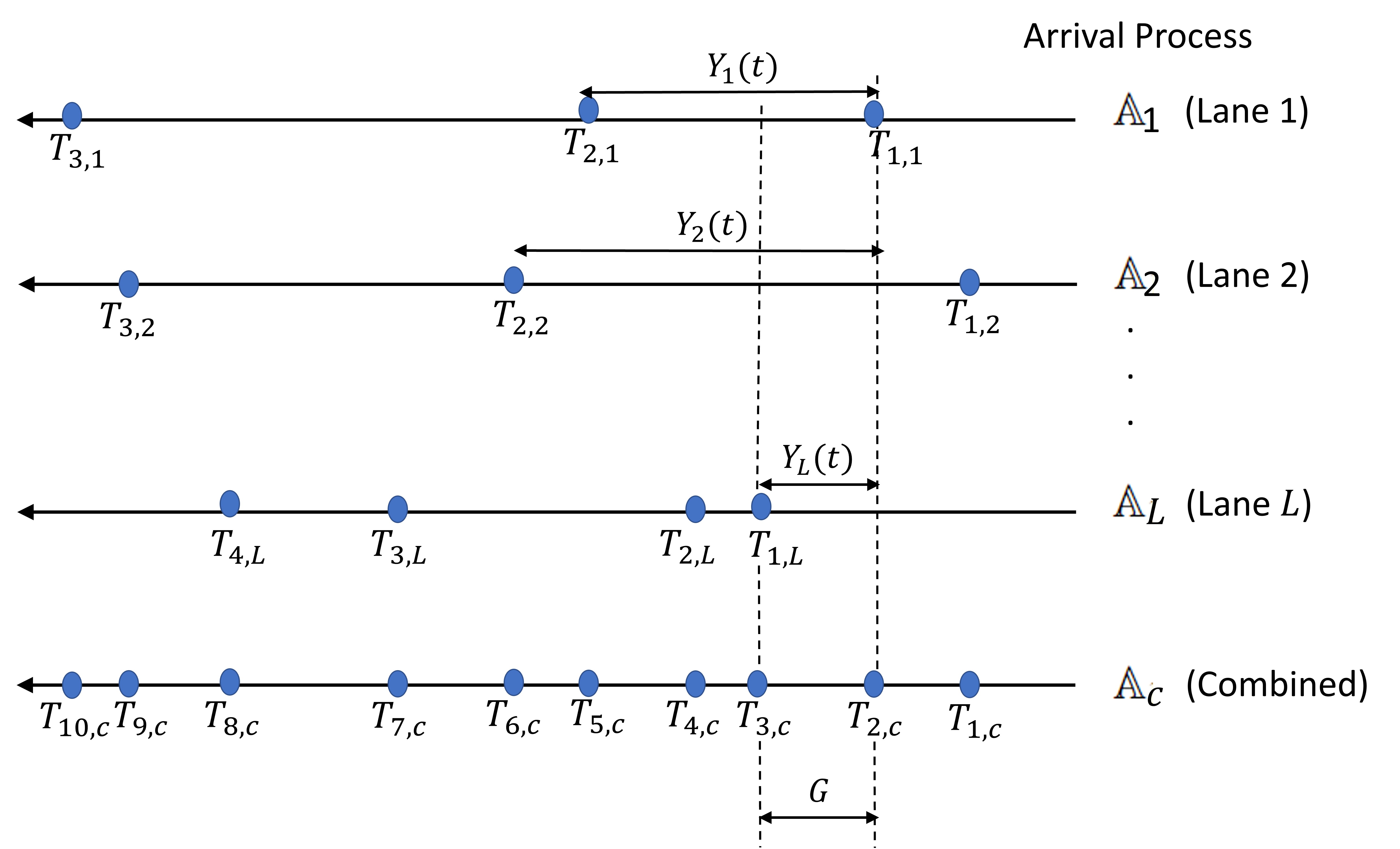}
    \caption{Gaps in combined arrival process}
    \label{fig:gap_appendix}
\end{figure}

The pdf of gaps, $f_G(g)$ given by Equation \ref{eq:pdf_gap} can be derived using an alternative approach. The method explained in Section \ref{sec:dist_gap} first finds the pdf of the remaining time of gaps in the combined process $(f_{Y_c}(y))$ and then from that pdf of gaps $(f_G(g))$ is determined with the help of Equation \ref{eq:pdf_rem}. But here $f_G(g)$ is determined directly. Consider Figure \ref{fig:gap_appendix} and note that in this figure $t = T_{2,c}$ (where $t$ is the time when one vehicle has just arrived at Section $\mathbb{S}$) whereas in Figure \ref{fig:deriv}, $t$ was between $T_{2,c}$ and $T_{3,c}$ (that is, $t$ is any time between arrival two successive vehicles at Section $\mathbb{S}$). So, the gap, $G$, in the combined arrival process can be written as,
\begin{center}
    $G = \min(Y_1(t),Y_2(t),\cdots,Y_L(t))$
\end{center}
Note that the notation used are the same as introduced in Section \ref{sec:dist_gap}.

If $F_G(g)$ is the cumulative distribution function of gaps then,
\begin{equation*}
    \begin{split}
    F_G(g) &= P(G \le g) = P(\text{at least one of $Y_1(t), Y_2(t), \cdots Y_L(t)$ } \le g)\\
           &= 1 - P(Y_1(t)>g, Y_2(t)>g,\cdots Y_L(t)>g)\\
    \end{split}
\end{equation*}
While finding gap at time $t$, one of the $Y_1(t), Y_2(t),... Y_L(t)$ is the inter-arrival time of one of the processes and the rest of them are the remaining times. The probability of $Y_j(t)$ being the inter-arrival time is proportional to the flow in $j^\text{th}$ process.
\begin{equation*}
 P(Y_j \sim F_{H_j}(g)) = P(\text{$Y_j$ is distributed as $F_{H_j}(g)$})  = \frac{\frac{1}{\mu_j}}{\sum_{s=1}^{L} \frac{1}{\mu_j}}
\end{equation*}
Therefore,
\begin{equation*}
    \begin{split}
    1-F_G(g)&=P(Y_1(t)>g, Y_2(t)>g,\cdots, Y_L(t)>g) \\
          &= P(Y_1(t)>g, Y_2(t)>g,\cdots,Y_L(t)>g | Y_1 \sim F_{H_1}(g)).P(Y_1\sim F_{H_1}(g))+\\
          &\quad \ P(Y_1(t)>g, Y_2(t)>g,\cdots,Y_L(t)>g| Y_2 \sim F_{H_2}(g)).P(Y_2\sim F_{H_2}(g))+\\
          &\quad \ ...\ +\\
          &\quad \ P(Y_1(t)>g, Y_2(t)>g,\cdots,Y_L(t)>g | Y_L \sim F_{H_L}(g)).P(Y_L\sim F_{H_L}(g))
    \end{split}
\end{equation*}
From Equation \ref{eq:cdf_rem_head} and \ref{eq:pdf_rem}, and for large $t$, the above expression can be written as,
\begin{equation*}
    \begin{split}
    1-F_G(g) \ 
    &= (1-F_{H_1}(g))(1-F_{Y_2}(g))\cdots(1-F_{Y_L}(g)).\left(\frac{\frac{1}{\mu_1}}{\sum_{s=1}^{L} \frac{1}{\mu_s}}\right)+\\
    &\quad \ (1-F_{Y_1}(g))(1-F_{H_2}(g))\cdots(1-F_{Y_L}(g)).\left(\frac{\frac{1}{\mu_2}}{\sum_{s=1}^{L} \frac{1}{\mu_s}}\right)+\\
    &\quad \ \cdots \ +\\
    &\quad \ (1-F_{Y_1}(g))(1-F_{Y_2}(g))\cdots(1-F_{Y_L}(g)).\left(\frac{\frac{1}{\mu_L}}{\sum_{s=1}^{L} \frac{1}{\mu_s}}\right)\\
    &\quad\\
    &= \sum_{j=1}^L \left(1-F_{H_j}(g)\right) \left(\prod_{k = 1, k\neq j}^L \left(1-F_{Y_k}(g) \right)\right) \left(\frac{\frac{1}{\mu_j}}{\sum_{s=1}^L \frac{1}{\mu_s}}\right)
    \end{split}
\end{equation*}
Therefore,
\begin{equation*}
    \begin{split}
    F_G(g) &= 1-\sum_{j=1}^L \left(1-F_{H_j}(g)\right) \left(\prod_{k = 1, k\neq j}^L \left(1-F_{Y_k}(g) \right)\right) \left(\frac{\frac{1}{\mu_j}}{\sum_{s=1}^L \frac{1}{\mu_s}}\right)\\
         &= 1- \sum_{j=1}^L \left(1-F_{H_j}(g)\right) \left(\prod_{k = 1, k\neq j}^L \left(1-\frac{1}{\mu_k} \int_{0}^g \left(1-F_{H_k}(h)\right) dh \right)\right) \left(\frac{\frac{1}{\mu_j}}{\sum_{s=1}^L \frac{1}{\mu_s}}\right)
    \end{split}
\end{equation*}
The density function of the final inter-arrival time is:
\begin{equation*} 
    \begin{split}
    f_G(g) &= \frac{d}{dg} F_G(g)\\
           &= \sum_{j=1}^{L} \Bigg[  \left(\frac{\frac{1}{\mu_j}}{\sum_{s=1}^{L}\frac{1}{\mu_s}}\right) \Bigg(f_{H_j}(g) \left(\prod_{k = 1, k\neq j}^{L} (1-F_{Y_k}(g))\right) +(1-F_{H_j}(g)) \nonumber\\
           & \quad \quad \quad \quad \quad \quad \quad \quad \quad \quad  \left(\sum_{k=1,k\neq j}^L (f_{Y_k}(g)) \prod_{m = 1,m \neq j,k}^L (1-F_{Y_m}(g))\right)  \Bigg)  \Bigg]
    \end{split}
\end{equation*}
The expression for $f_G(g)$ obtained here is the same as the one mentioned in Equation \ref{eq:pdf_gap}.

\section{Example for non-uniqueness of parameters} \label{app:non_unique}
\begin{figure}[H]
    \centering
    \includegraphics[width=0.7\textwidth]{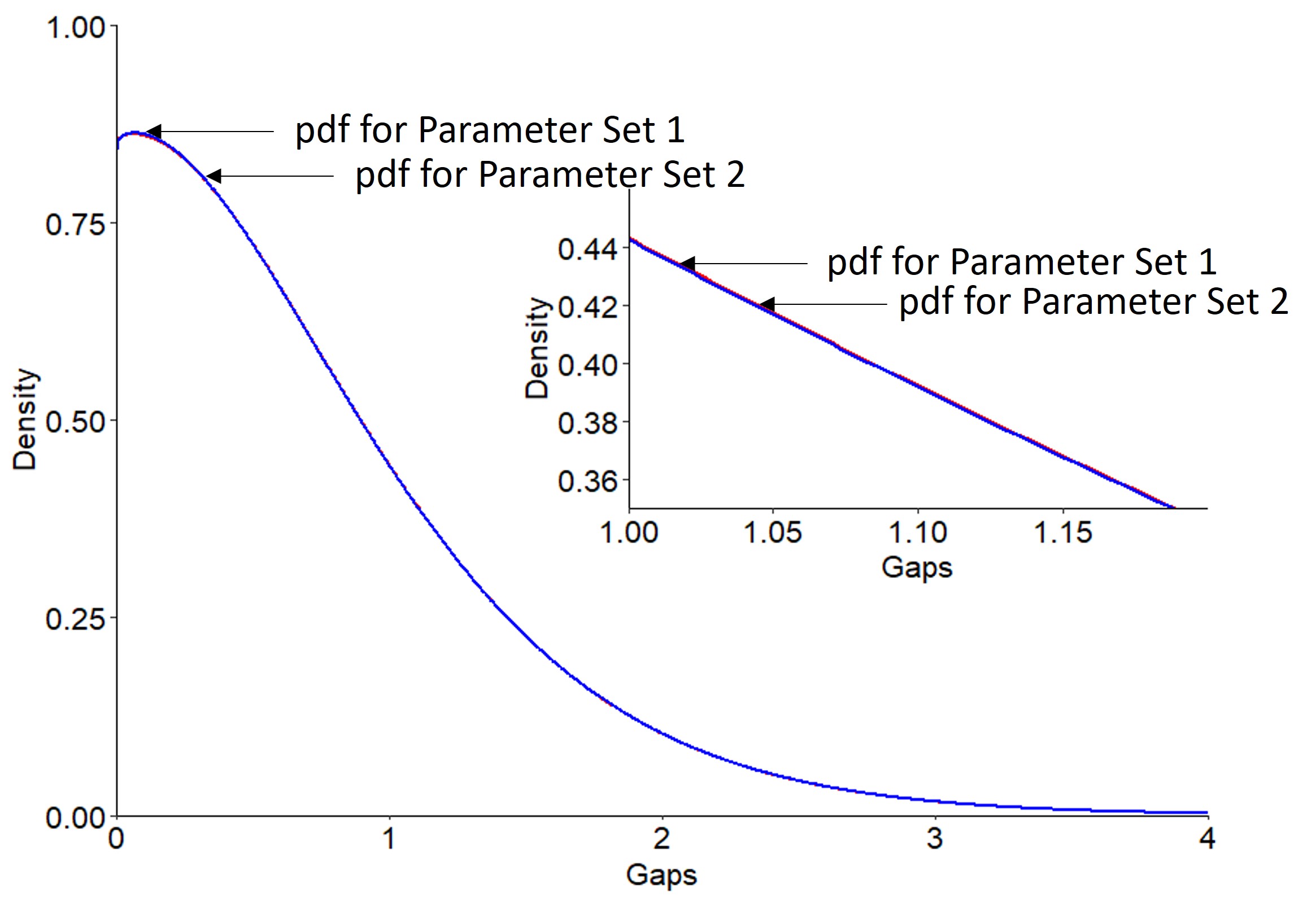}
    \caption{pdf of gaps, $f_G(g)$, for two parameter sets; inset is a magnified plot of the pdf}
    \label{fig:non_unique}
\end{figure}

As discussed in Section \ref{sec:est_from_gap}, there may be many sets of values of parameters that give rise to the indistinguishable gap distributions. A formal proof for the previous assertion remained elusive and is probably unnecessary for this study. Instead, the claim of indistinguishability is demonstrated through an illustrative example. Figure \ref{fig:non_unique} plots the pdf of gap distribution for two sets of parameter values. For Parameter Set 1, the $k_j$ and $\lambda_j$ (where, $j \in \{1,2,3,4$\}) values are 1.5, 2.0, 2.6, 3.6 and 0.5, 0.8, 1.3, 0.1, respectively and for Parameter Set 2, the $k_j$ and $\lambda_j$ (where, $j \in \{1,2,3\}$) values are 1.3, 2.6, 2.26 and 0.45, 1.05, 1.15, respectively. Even though the pdf of the gap is plotted for two sets of parameters only one line can be seen in the figure because the differences are imperceptible even in the magnified inset.
\vspace{-12pt}

\section{Detailed result of estimates in disorderly streams} \label{app:table}
Table \ref{tab:est_disorder_gap_all_L} shows the parameter estimates for all the values of $L$ with their maximum log-likelihood values. Note, although the MLE values are the same for each $L$, the AIC is best for $L = 1$ as it has the least number of parameters. Further, for $L\ge 2$, only one value of the $\lambda_j$ parameters is meaningful, the rest are all insignificant. This meaningful $\lambda_j$ (and $k_j$) parameter value was the same for all $L$. This further indicates that $L=1$ is a reasonable estimate.

\begin{table}[H]
\centering
\caption{Detailed estimates of parameters for disorderly streams}
\label{tab:est_disorder_gap_all_L}
\makebox[1 \textwidth][c]{
\resizebox{1.1 \textwidth}{!}{
\begin{tabular}{ccccccccc}
    \hline
    \multirow{4}{*}{$L$}& \multirow{4}{*}{$j$} &\multicolumn{3}{c}{Kanpur} & & \multicolumn{3}{c}{Chennai} \\
    \cline{3-5} \cline{7-9}
     &  &\multirow{3}{*}{$\hat{k}_j$ (t-value)} & \multirow{3}{*}{$\hat{\lambda}_j$ (t-value)} & Maximum& & \multirow{4}{*}{$\hat{k}_j$ (t-value)}  & \multirow{3}{*}{$\hat{\lambda}_j$ (t-value)} & Maximum \\
     &    &  &  & log-  & &  &   & log-      \\
     &    &  &  & likelihood  & &  &   & likelihood      \\
    \hline
    1                  & 1    & \textbf{0.820} (66.6)   & \textbf{0.482}     (49.4)   & -10073.9  & & \textbf{1.169} (82.2) & \textbf{1.414}    (66.3)   & -8612.1        \\
\hline
\multirow{2}{*}{2} & 1    & \textbf{0.820}  (69.4) & \textbf{0.482}     (55.4)   & \multirow{2}{*}{-10073.9} && 2.628  (NaN)    & 3.61e-6  (3.24e-5) & \multirow{2}{*}{-8612.1}\\
                   & 2    & 8.092  (NaN)    & 4.23e-4  (3.6e-2)    &         &   & \textbf{1.168}  (75.2) & \textbf{1.414}     (30.1)   &        \\
\hline
\multirow{3}{*}{3} & 1    & 3.460  (NaN)    & 6.69e-6  (1.14e-4) & \multirow{3}{*}{-10073.9}& &\textbf{1.168}  (65.6) & \textbf{1.413}     (18.2)   & \multirow{3}{*}{-8612.1} \\
                   & 2    & 2.425  (NaN)    & 2.37e-6  (1.66e-4) &  &   & 2.375  (5.8)  & 2.49e-6  (1.96e-4) &           \\
                   & 3    & \textbf{0.820}  (66.6) & \textbf{0.482}     (49.5)   &  &  & 2.829  (NaN)    & 8.62e-7  (8.69e-6) &           \\
\hline
\multirow{4}{*}{4} & 1    & 3.464  (21.8) & 7.91e-7  (1.07e-5) & \multirow{4}{*}{-10073.9}& &\textbf{1.176}  (59.3 )& \textbf{1.421}     (17.4)   & \multirow{4}{*}{-8612.1} \\
                   & 2    & 2.647  (3.5)  & 1.01e-5 (1.79e-4)  &  &  & 2.447 (5.0)  & 3.16e-4  (2.36e-3) &                         \\
                   & 3    & \textbf{0.820}  (49.6) & \textbf{0.482} (13.4)       &  &   & 3.102  (NaN )   & 2.93e-6  (2.26e-5) &                        \\
                   & 4    & 2.340  (1.7)  & 7.06e-5  (1.00e-3)    &    &   & 4.129  (NaN)    & 8.24e-6  (4.77e-5) &                      \\
\hline
\multirow{5}{*}{5} & 1    & 3.510  (NaN)    & 1.53e-4  (3.00e-3)    & \multirow{5}{*}{-10073.9} && \textbf{1.169}  (56.758) & \textbf{1.411}     (14.1)   & \multirow{5}{*}{-8612.1}\\
                   & 2    & \textbf{2.750}  (15.1) & 1.40e-5  (2.44e-4)  &   &    & 3.561  (8.0)  & 7.80e-7  (4.08e-6) &                       \\
                   & 3    & \textbf{0.820}  (47.2) & \textbf{0.482}     (12.2)   &    &    & 2.688  (5.6)  & 2.72e-6  (1.86e-5) &                    \\
                   & 4    & 1.790  (NaN)    & 1.20e-6  (8.69e-5) &    &     & 4.871  (7.8 ) & 4.56e-5  (1.53e-4) &                    \\
                   & 5    &2.099  (NaN)     & 7.45e-6  (1.71e-4) &    &    & 4.316  (NaN)    & 0.003     (1.60e-2) &     \\
    \hline
\end{tabular}
}
}
\end{table}

\end{appendix}
\end{document}